# FOCUSED ION BEAM FABRICATION AND IBIC CHARACTERIZATION OF A DIAMOND DETECTOR WITH BURIED ELECTRODES

P. Olivero[1], J. Forneris[1], M. Jakšić[2], Ž. Pastuović[2], F. Picollo[1], N. Skukan[2], E. Vittone[1†]

[1]Experimental Physics Department- NIS Excellence Centre, University of Torino and INFN-sez. Torino, via P.Giuria 1, 10125 Torino, Italy

[2] Ruđer Bošković Institute, Bijenicka 54, P.O. Box 180, 10002 Zagreb, Croatia

**ABSTRACT**

This paper reports on the fabrication and characterization of a high purity monocrystalline diamond detector with buried electrodes realized by the selective damage induced by a focused 6 MeV carbon ion beam scanned over a pattern defined at the micrometric scale. A suitable variable-thickness mask was deposited on the diamond surface in order to modulate the penetration depth of the ions and to shallow the damage profile toward the surface. After the irradiation, the sample was annealed at high temperature in order to promote the conversion to the graphitic phase of the end-of range regions which experienced an ion-induced damage exceeding the damage threshold, while recovering the sub-threshold damaged regions to the highly resistive diamond phase. This process provided conductive graphitic electrodes embedded in the insulating diamond matrix; the presence of the variable-thickness mask made the terminations of the channels emerging at the diamond surface and available to be connected to an external electronic circuit. In order to evaluate the quality of this novel microfabrication procedure based on direct ion writing, we performed frontal Ion Beam Induced Charge (IBIC) measurements by raster scanning focused MeV ion beams onto the diamond surface. Charge collection efficiency (CCE) maps were measured at different bias voltages. The interpretation of such maps was based on the Shockley-Ramo-Gunn formalism.

---

[†] Corresponding author: ettore.vittone@unito.it

## 1. INTRODUCTION

Diamond has been systematically studied since 1940s to make solid state ionization chambers for ionization radiation detection. In fact, its extreme properties make this material appealing for many applications, ranging from high energy physics experiments to radiotherapy dosimetry, neutron spectroscopy and monitoring of the activity in corrosive nuclear waste solutions [1].

The advances in the CVD growth technology recently made available to the market "detector grade" diamond samples with superior features, from both a structural (high crystal perfection, low impurity density) and an electronic (high carrier mobilities and lifetimes) point of view [2]. However, the full exploitation of the potential of diamond as a material for ionizing radiation detectors requires the availability of techniques that allow the modification of the structural and electrical properties of single crystals at a micrometric scale.

In this paper we show that focused MeV ion beams can effectively contribute to both the fabrication and the characterization of a diamond detector with active regions defined with micrometer resolution.

For what concerns the fabrication, we have realized buried conductive electrodes by exploiting the phase transformation of selected regions irradiated with C ions from diamond to a graphitic phase. Such phase transition only occurs in regions where the damage density overcomes the so-called "graphitization threshold" (i.e. the vacancy density above which the damaged diamond structure permanently converts to graphite upon thermal annealing), thus resulting in the formation of well-defined graphitic layers at the end-of-range of the implanted ions [3, 4]. The electrical contacts were realized with a novel technique based on the employment of variable-thickness metallic masks that allowed the modulation of the depth of the buried electrodes, thus allowing their electrical contacting with the surface at their endpoints [5, 6].

For what concerns the characterization, we employed the IBIC technique to map the charge collection efficiency of the ion-microfabricated diamond detector. This technique allows a direct imaging of the active regions and a deep insight into the mechanisms occurring in induced-charge pulse formation [7].

## 2. EXPERIMENTAL

The sample used in this work is a synthetic single crystal diamond produced by ElementSix Ltd. with Chemical Vapor Deposition (CVD) process [2]. The crystal is 2.0×2.0×0.5 mm$^3$ in size and is classified as type IIa "Electronic Grade" due to its high purity (N and B concentration: <5 ppb) and lattice perfection. The crystal consists of a single growth sector, oriented in the {100} direction on the two opposite large faces, which are polished down to a mean roughness $R_a$ < 5 nm.

Variable-thickness structures were realized on the diamond surface by thermal evaporation of copper through a patterned mask consisting of an array of square holes (250×90 µm$^2$) and kept at a distance of about 300 µm from the sample surface. This arrangement (as shown schematically in Fig. 1) allowed the deposition of a rectangular area (300×105 µm$^2$) of Cu with an average thickness of 5 µm in the central region and with smooth and slowly degrading edges at each of the four corners.

The MeV ion implantation was performed at the Laboratory for Ion Beam Interaction of the Ruđer Bošković Institute using a 6 MeV C$^{3+}$ focused beam raster scanned along the shorter sides of the rectangular area, i.e. along two linear paths (~30 µm wide and ~150 µm long) terminating at the corners 1-2 and 3-4 (Fig. 1, labels C), where the masking metal layers were located. Similarly, two additional linear scans (~15 µm wide, and ~130 µm and ~165 µm long respectively) of the focused ion beam were carried out perpendicularly to the previous ones (Fig. 1, labels B). The resulting pattern, drawn in the scheme in Fig. 1 and shown in the micro-photograph in Fig. 2, was realized in about 30 min, using a beam current of ~2 nA in order to achieve an implantation fluence of $4\cdot10^{16}$ ions cm$^{-2}$. Such value was chosen in order to be certain that the induced vacancy density would exceed the graphitization threshold. In fact, the vacancy profile (Fig. 3) can be roughly evaluated by multiplying the ion fluence with the vacancy linear density profile per ion calculated with SRIM-2008.04 Monte Carlo Simulation code [8], assuming a displacement energy value of 50 eV [9]. Although such damage profile has to be considered as a rough estimation, since it results from a linearly cumulative effect of ion damage (i.e. neglecting any saturation effects occurring at high damage levels such as self-annealing and vacancies interactions), in this context it has been considered sufficient to estimate the dimensions of the buried region which, after thermal annealing, undergoes a conversion from the amorphous damaged diamond phase to a graphitic phase.

If a graphitization threshold of $9\cdot10^{22}$ vacancies cm$^{-3}$ [5] is assumed, such region can be estimated to be approximately 300 nm in thickness, and is located at an average depth of 2.7 μm, as evidenced by the vertically shaded area in Fig. 3. When ions pass through a thin Cu layer before entering the diamond, the highly damaged end-of-range layer becomes more shallow, as shown by the horizontally shaded area in the same figure. As a consequence, the scan of the ion beam through the smooth and degrading profile of the Cu layers makes these highly damaged regions emerging at the diamond surface, as shown in the schematics of Fig. 1.

After ion implantation, the Cu masks were removed from the sample surface, and the diamond was annealed in Ar atmosphere at 950 °C for 2 hours. Fig. 2 shows the transmission micro-photograph of the sample after annealing: the buried graphitic channels are clearly visible, due to their strong opacity with respect to the transparent diamond matrix.

The connection of the buried channels to an external electronic circuit was accomplished by depositing thin Ag layers at the emerging terminals (labeled as 1-4 in Fig. 1). The electrical continuity of the buried channels with the surface electrodes and the resistance of the channels were evaluated by two-terminal current-voltage (I-V) measurements between contacts 1 and 2, as well as between contacts 3 and 4 (see Fig. 1). In both cases the I-V characteristics were linear in the range from -1 V to +1 V with a slope (i.e. a conductance) of 0.8 mS. Assuming a graphitic layer thickness of 300 nm, a width of 32 μm and a length of 150 μm, the layer resistivity can be estimated as $8 \cdot 10^{-3}$ Ω cm, to be compared with the resistivity of HOPG graphite of $10^{-5}$ Ω cm along the basal planes [10] and to the resistivity of pristine diamond of $>10^9$ Ω cm. Moreover, the negligible current of < 1 nA measured when a bias voltage of 100 V was applied between contacts 1 (shorted with 2) and 4 (shorted with 3), demonstrates that the two buried electrodes are insulated from each other, as expected from being embedded in a strongly insulating diamond matrix.

IBIC measurements were carried out using 2 MeV protons focused down to a 2 μm diameter spot and raster-scanned onto the rectangular area of 300×300 μm$^2$ surrounding the buried channels. Referring to Fig. 1, contacts 3-4 were grounded, while contacts 1-2 were biased at variable voltages and connected to a charge-sensitive electronic chain and to the acquisition system described in [11]. An additional grounded electrode was realized on the back side of the diamond, i.e. on the surface opposite to the irradiated side.

The calibration of the electronic chain was performed by using a Si surface barrier detector and a precision pulse generator, in order to relate pulse heights provided by the reference Si detector, for which 100% charge collection efficiency was assumed, with those from diamond. Assuming an average energy of 3.62 eV and 13.2 eV to create electron/hole pairs in Si and diamond, respectively [12], the spectral sensitivity of the IBIC set-up was 630 electrons/channel, corresponding to about 0.4% charge collection efficiency for 2 MeV protons in diamond.

## 3. Results

Fig. 4 shows IBIC maps collected at different bias voltages. The charge collection efficiency (CCE) is encoded in the colour scale, which represents the median of the IBIC pulse distribution for each pixel. The schematics of the two electrodes are superimposed to the IBIC maps and profiles to show their relative positions.

The circular regions close to the endpoints of the vertical channels show unexpectedly high collection efficiencies, which evidence the presence of a high electric field probably due to non-perfect ohmic contacts between the emerging graphitic channels and the top silver electrodes.

At low bias voltages, the most efficient region is located around the grounded channel. As the bias voltage increases, the high efficiency region widens and, at the maximum bias voltage, an almost complete charge collection is achieved in the region between the buried electrodes.

Moreover, the CCE profiles (Fig. 4D) evaluated for different bias voltages along the region crossing perpendicularly the buried electrodes at x=130-148 μm exhibit a maximum CCE at the grounded electrode, whereas a smaller CCE can be observed around the sensitive electrode. This effect can be attributed to a dominant contribution of electrons with respect to holes to the formation of the IBIC signal.

Although a detailed analysis of these IBIC maps is beyond the scope of the present paper, a preliminary two-dimensional model of the mechanism underlying the charge collection can provide some insight that supports the observations. The model was developed on the basis of the theory outlined in previous papers [13, 14, 15].

One of the most fundamental conclusions of the Shockley-Ramo-Gunn's theory is that the induced charge signal Q is proportional to the difference in the weighting potentials between any two positions of the moving charge [16]:

$$1)\quad Q \propto \left[ \left.\frac{\partial \psi}{\partial V}\right|_r - \left.\frac{\partial \psi}{\partial V}\right|_{r_0} \right]$$

where ψ is the actual potential, V is the applied bias voltage, $\frac{\partial \psi}{\partial V}$ is the Gunn's weighting potential (GWP); $r_0$ and r are the initial and final point of the trajectory of a charge carrier moving in the electric field region, respectively. Hence, according to eq. 1), as long as a non-zero weighting potential exists inside the active region where the charge movement occurs, a charge signal will be induced.

Fig. 5a shows the contour plot of the GWP with 100 equally spaced levels ranging from 0 to 1 as evaluated by solving numerically the Laplace's equation with the Finite Element Method implemented in the commercial software COMSOL Multiphysics 3.5a® [17]. The integration domain corresponds to the plane (y-z) crossing the two buried electrodes between x=130 μm and 148 μm (see maps in Fig. 4).

In our experiments, charge carriers are generated by ionization within 25 μm from the surface, corresponding to the range of 2 MeV protons in diamond. Within such charge generation profile, the maximum variability of the GWP occurs around and between the two buried electrodes, where a very high electric field occurs. As a consequence, the carriers drifted in the high electric field with their saturation velocity of the order of $10^{11}$ μm s$^{-1}$ in diamond across this region, and the probability for a carrier of incurring into recombination mechanism during the very short drift time is so low that almost all the generated charge contributes to the IBIC signal.

Outside the region between the two buried electrodes, the GWP is not null, although its variability is much less pronounced. This fact accounts for the long tails in the CCE profiles occurring externally to the two electrodes mainly at high bias voltages.

For a more detailed analysis of the CCE profile shown in Fig. 4D, we have calculated the CCE map relevant to the same domain shown in Fig. 5a by solving the adjoint continuity equations for electrons and holes, as reported in [13, 15]. In this numerical calculation, whose results are reported in Fig. 5b, the applied bias potential was 10 V, the values of carrier mobilities were extracted from [2] and the electron and hole lifetimes were set to 20 ns and 0.5 ns, respectively. As expected, the longer drift length of electrons makes the region around the grounded electrode more efficient for the IBIC pulse formation.

Fig. 6 shows the CCE profile obtained with numerical simulations by convoluting the Bragg's generation profile relevant to 2 MeV protons in diamond with the CCE map of Fig. 5b. Although far from being perfectly adherent to the experimental profile, this result corroborates the above mentioned qualitative interpretation. In particular, we conclude that the contribution to the CCE is mainly due to electrons: their drift from the grounded to the sensitive electrode provides a linear behaviour of the CCE profile that is characteristic of a situation in which a constant (saturation) drift velocity occurs [14]. The asymmetry of the CCE profile accounts for the asymmetry of the electric field profile and, as a consequence, of the GWP, as shown in Fig. 5a. Due to the shorter lifetime, the contribution of holes is essentially restricted to regions close to the sensitive electrode (anode).

## 4. Conclusions

In this paper we report on the direct writing of buried conductive channels in a monocrystalline "electronic grade" diamond sample using a focused MeV ion beam. The technique is based on the phase transformation of diamond which evolves into a $sp^2$ rich (i.e. graphitic) phase because of the lattice damage due to ion irradiation. Although this process has been used by several authors for diamond micromachining, we present in this paper a method based on the use of a variable-thickness mask to produce buried contacts with endpoints that are emerging at the surface for a suitable electrical connection of the conductive channels to the external electronic circuit.

This process offers the advantage of achieving the fabrication of emerging channels during the ion implantation, without further fabrication processes such as high-power laser graphitization [18], multiple-energy implantation [19], or electric-breakdown-induced graphitization [4]. Another advantage consists in the use of a single evaporation mask both to fabricate the variable-thickness layer, and to deposit the Ag electrodes onto the emerging endpoints of the graphitic channels after thermal annealing. The conductivity of these graphitic channels was proven to be more than ten orders of magnitude higher than what was measured before ion irradiation.

In order to valuate the adequacy of this procedure to fabricate a diamond detector for ionizing radiation, we performed an IBIC characterization using a 2 MeV $H^+$ ion microbeam. The CCE maps show excellent performances in regions around and in between the buried electrodes. For high bias voltage, the CCE maintains values around 70% at distances of the order of 100 μm from the electrodes. A preliminary analysis of the IBIC maps was conducted on the basis of the Shockley-Ramo-Gunn's theory. In spite of the simplicity of model, which neglects any space charge (polarization) effect, assumes constant carrier lifetimes throughout the entire volume and is limited to the calculation of a two dimensional profile along a section perpendicular to the buried electrodes, the simulation provides evidence of the dominant role played by electrons in the IBIC pulse formation mechanism.

These charge collection performances make this system promising for the development of a novel generation of three-dimensional diamond detectors based on inter-digitated buried electrodes.

Acknowledgements

The work of P. Olivero is supported by the "Accademia Nazionale dei Lincei – Compagnia di San Paolo" Nanotechnology grant, which is gratefully acknowledged.

**Figure captions**

Fig. 1. Three-dimensional scheme (not in scale) of the conductive channels (C) and electrodes (B) buried in monocrystalline diamond (A). The Ag metal layers deposited onto the emerging terminals are labelled with as 1-4: the inset shows the variable-thickness mask of the Cu layer evaporated before ion implantation.

Fig. 2. Optical photo-micrograph in transmission of the sample after thermal annealing; the heavily damaged regions are clearly visible since they are optically opaque with respect to the transparent diamond matrix.

Fig. 3. Vacancy density profile in diamond for 6 MeV C implantation at fluence of $4 \cdot 10^{16}$ ions cm$^{-2}$ in absence (profile on the right) and in presence (profile on the left) of a 2-μm-thick Cu layer deposited onto the diamond surface. The dashed areas correspond to the regions where the vacancy density exceeds the graphitization threshold ($9 \cdot 10^{22}$ vacancies cm$^{-3}$ [5]).

Fig. 4. IBIC maps collected at bias voltages of +10 V(A), +20 V(B), +40 V(C). The electrodes geometry is schematically superimposed to the maps and profiles to show their relative positions. (D) IBIC profiles along the y-direction calculated by projecting the average CCE of the region between x=130 μm and x=148 μm on the y-axis. The two rectangles represent the positions of the electrodes.

Fig. 5. a) Contour plot of the Gunn's weighting potential along the y-z plane with 100 equally spaced levels ranging from 0 to 1 (at the sensitive electrode, GWP=1; at the grounded electrode GWP=0). b) Surface plot of the CCE across the same area.

Fig. 6. Total CCE profiles containing both electron and hole contribution calculated numerically by the convolution of the CCE map of Fig. 5 with the ionization profile of 2 MeV protons in diamond.

**Figures**

**Fig. 1**

**Fig. 2**

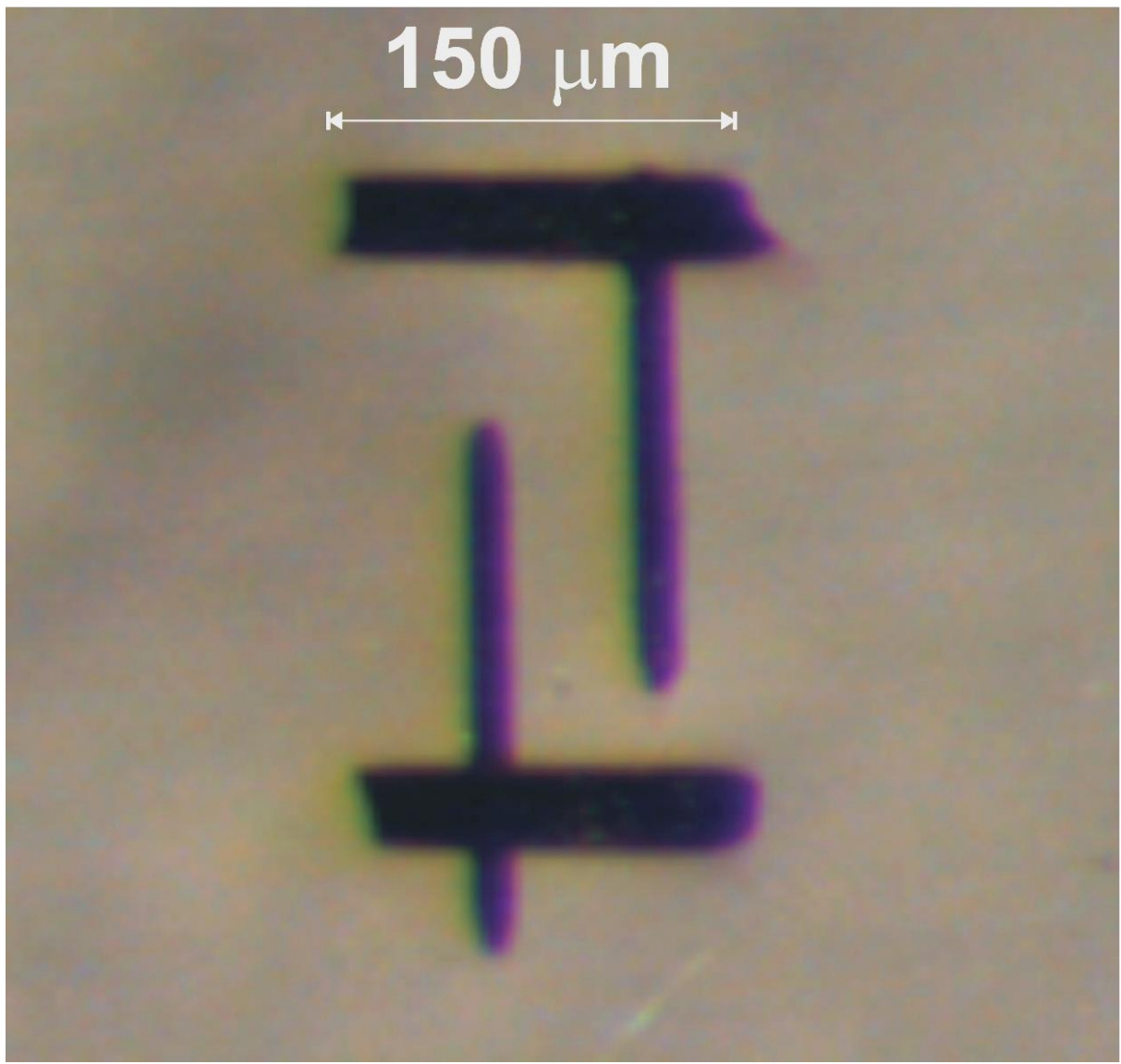

**Fig. 3**

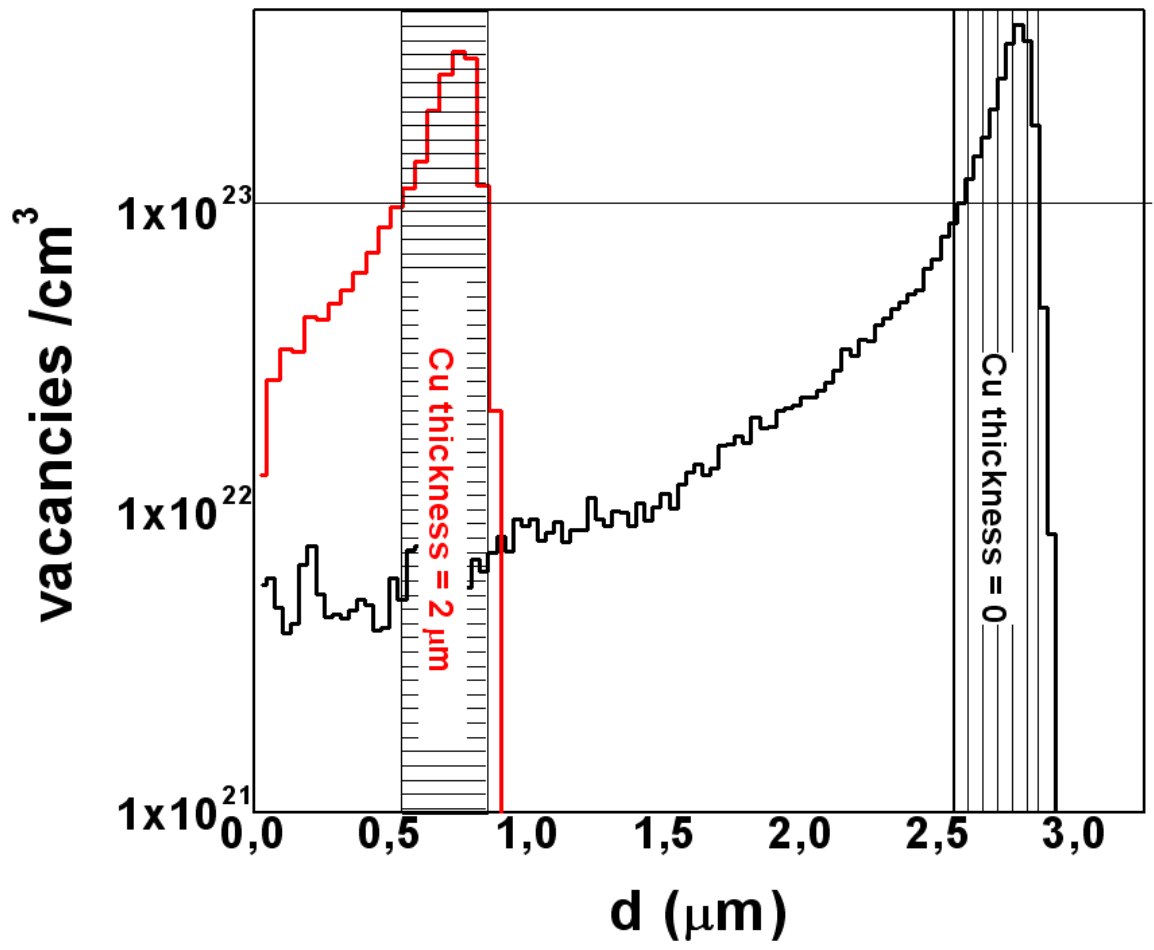

**Fig. 4**

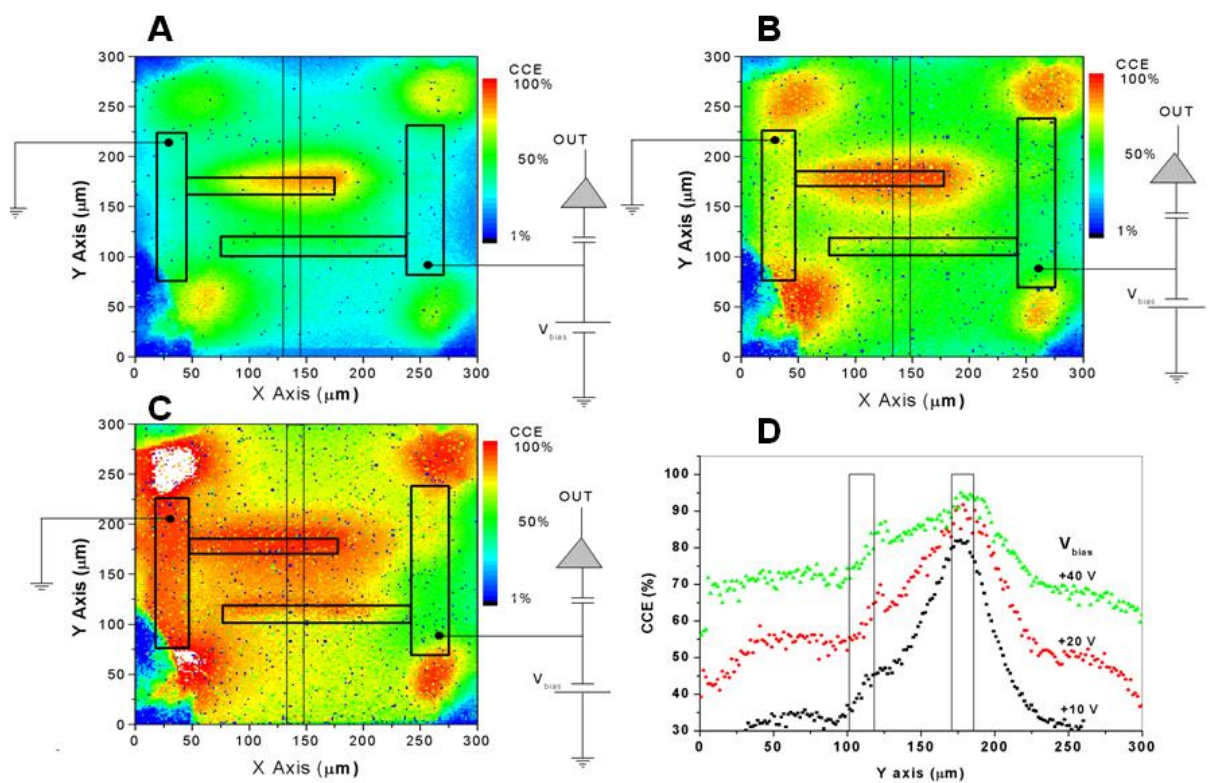

**Fig. 5**

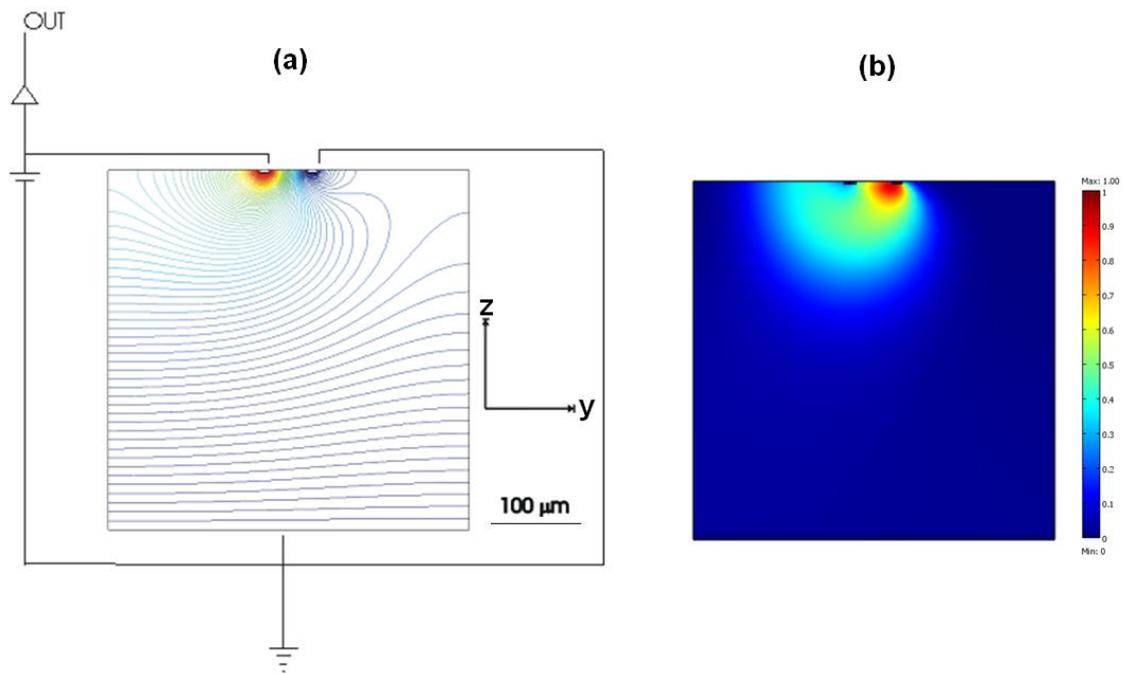

**Fig. 6**

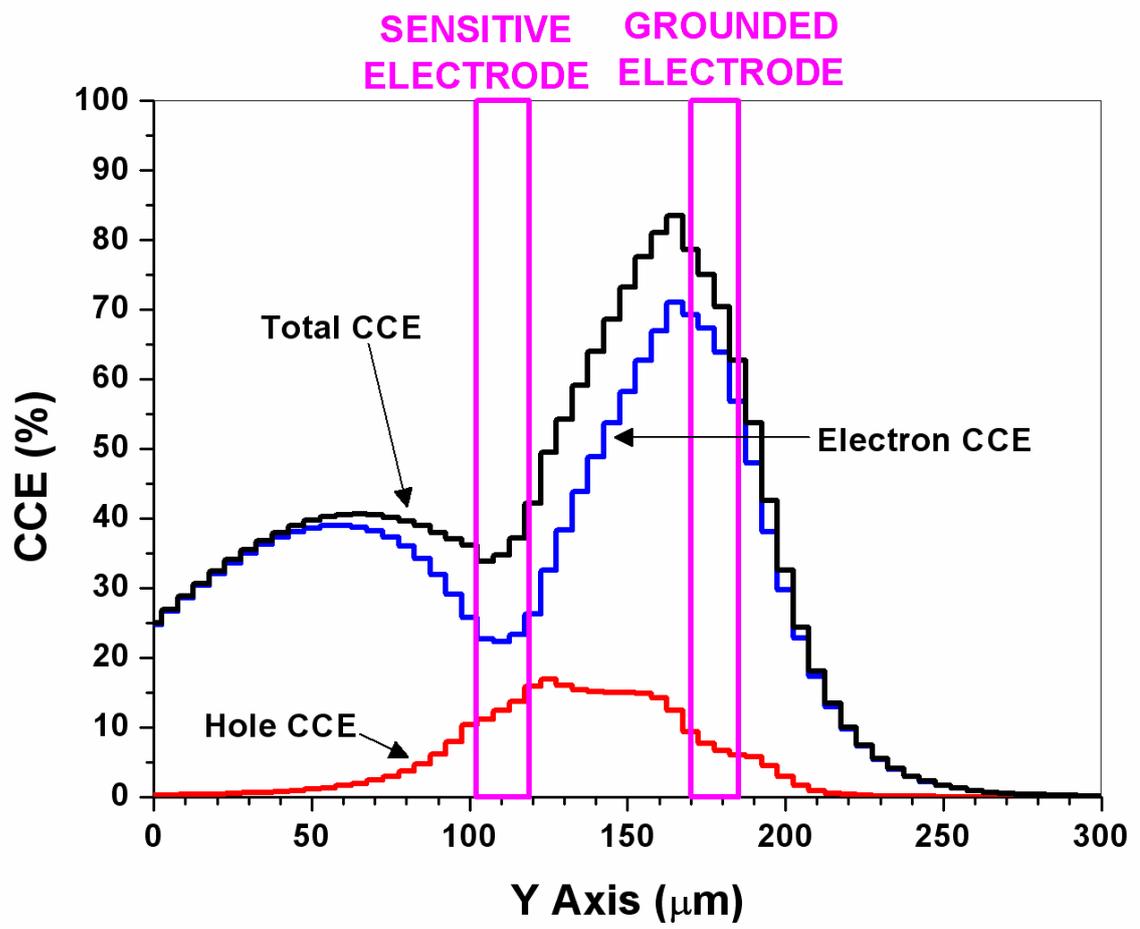